\documentclass[a4paper]{amsart}
\makeatletter
\newif\if@restonecol
\makeatother

\usepackage{tikz}
\usetikzlibrary{shapes,arrows}
\usepackage{url}
\usepackage{amsmath,bm,times}
\usepackage{multirow}
\usepackage{multicol}
\usepackage{algorithm2e}
\usepackage{graphicx}
\usepackage{tikz}
\usepackage{pgfplots}
\usepackage{geometry}
\newtheorem{thm}{Theorem}

\theoremstyle{definition}
\newtheorem{defn}[thm]{Definition}
\theoremstyle{remark}

\begin{document}

\title[NTRFinder: A Software Tool to Find Nested Tandem Repeats]{NTRFinder: A Software Tool to Find Nested Tandem Repeats}


\author{A.\ A.\ Matroud}
\address{Allan Wilson Centre for Molecular Ecology and Evolution\\\ Massey University\\ Private Bag 11 222\\ Palmerston North 4442\\ New Zealand}
\author{M.\ D.\ Hendy}
\address{Institute of Fundamental Sciences\\ Massey University \\Private Bag 11 222 \\ Palmerston North 4442 \\New Zealand}
\author{C.\ P.\ Tuffley}
\address{Institute of Fundamental Sciences\\ Massey University \\Private Bag 11 222 \\ Palmerston North 4442 \\New Zealand}

\begin{abstract}
We introduce the software tool {\tt NTRFinder} to find the complex
repetitive structure in DNA we call a nested tandem repeat (NTR). An
NTR is a recurrence of two or more distinct tandem motifs interspersed
with each other. We propose that nested tandem repeats can be used as
phylogenetic and population markers.

We have tested our algorithm on both real and simulated data, and
present some real nested tandem repeats of interest. We discuss how
the NTR found in the ribosomal DNA of taro ({\it Colocasia esculenta})
may assist in determining the cultivation prehistory of this ancient
staple food crop.

{\tt NTRFinder} can be downloaded from http://www.maths.otago.ac.nz/$\sim$aamatroud/ .
\end{abstract}

\maketitle

\section{Introduction}\label{int}

Genomic DNA has long been known to contain \emph{tandem repeats:}
repetitive structures in which many approximate copies of a common
segment (the \emph{motif}) appear consecutively. Several studies have
proposed different mechanisms for the occurrence of tandem
repeats~\cite{Weitzmann04041997, Wells02091996}, but their biological
role is not well understood.

Recently we have observed a more complex repetitive structure in the
ribosomal DNA of \emph{Colocasia esculenta} (taro), consisting of
multiple approximate copies of two distinct motifs interspersed with
one another.  We call such
structures \emph{nested tandem repeats} (NTRs), and the problem of
finding them in sequence data is the focus of this paper. Our
motivation is their potential use for studying populations: for
example, a preliminary analysis suggests that changes in the NTR
in taro have been occurring on a 1,000 year time scale, so a greater
understanding of this NTR offers the potential to date the early
agriculture of this ancient staple food crop.

The problem of locating tandem repeats is well known, as their
implication for neurological disorders \cite{huntington,Myotonic}, and
their use to infer evolutionary histories has urged some researchers to develop tools to find them. This has
resulted in a number of software tools, each of which has its own
strengths and limitations.  However, the existing tools were not
designed to find NTRs, and consequently do not generally find them.
In this paper, we present a new software tool, {\tt NTRFinder}, which
is designed to find these more complex repetitive structures.

We report here the algorithm on which {\tt NTRFinder} is based and
report some of the NTRs it has identified, including an even more
complex structure where copies of four distinct motifs are
interspersed.

\section{Background and Definitions}

\subsection{Sequences, edit operations and the edit distance}

A DNA sequence is a sequence of symbols from the nucleotide alphabet $\Sigma=\{{\tt A,C,G,T}\}$.
We define a DNA {\em segment} to be a string of contiguous DNA nucleotides and define a {\em site} to be a component in a segment.
For a DNA segment
$${\bf X}=x_1x_2\cdots x_n,$$
$x_i \in \Sigma$ is the nucleotide at the $i$-th site
and $|\mathbf{X}|=n$ is the length of $\mathbf{X}$.

Copying errors happen in DNA due to different external and internal
factors. These changes include substitution, insertion, deletion,
duplication, and contraction. We refer to these as \emph{edit
  operations}. By giving each type of edit operation some specific
weight, we can in principle find a series of edit operations which
transform segment $x$ to segment $y$, whose sum of weights is
minimal. We will refer to this sum as the \emph{edit distance}, and
denote it by $d(x,y)$. For the purposes of this paper, the edit
operations allowed in calculating the edit distance are single
nucleotide substitutions, and single nucleotide insertions or
deletions (indels), with each given weight 1.

\subsection{Classification of Tandem Repeats}

Many classifications of tandem repeat schemas have been introduced in the computational biology literature. We list some which are commonly used:
\begin{itemize}
\item{\bf (Exact) Tandem Repeats:} An {\em exact tandem repeat} (TR) is a sequence comprising two or more contiguous copies $\bf XX\cdots X$ of identical segments $\bf X$ (referred to as the {\em motif}).
\item{\bf $k-$Approximate Tandem Repeats:}  A {\em $k-$approximate tandem repeat} ($k-$TR) is a sequence comprising two or more contiguous copies ${\bf X}_1{\bf X}_2\cdots {\bf X}_n$ of similar segments, where each individual segment ${\bf X}_i$ is edit distance at most $k$ from a template segment $\bf X$.
\item{\bf Multiple Length Tandem Repeats:} A multiple length tandem repeat is a tandem repeat where each repeat copy is of the form ${\bf Xx}^n$, where $n$ is a constant larger than one and $d({\bf X,x})$ is greater than some threshold value $k$.
\end{itemize}

{\bf Examples:}
\begin{itemize}
\item{\bf TR:} \\   $\tt AGG~AGG~AGG~AGG~AGG$. The motif is $\tt AGG$.
\item{\bf$1-$TR:}\\ $\tt AGG~AG\underline C~A\underline TG~AGG~\underline CGG$. The motif is $\tt AGG$.
\item{\bf MLTR}:\\ $\tt GACCTTTGG~ACGGT~ACGGT~ACGGT$\\      $\tt GACCTTTGG~ACGGT~ACGGT~ACGGT$.~\\ The motifs are $\tt ACGGT$ and $\tt GACCTTTGG$, with $n=3$.

\end{itemize}

\subsection{Nested Tandem Repeats}
In this section we introduce a more complex repetitive structure,
the nested tandem repeat (NTR), also referred to as a {\em variable length
  tandem repeat} \cite{DBLP:conf/ismb/HauthJ02}.  Let $\mathbf X$ and
$\mathbf x$ be two segments (typically of different lengths) from the
alphabet $\Sigma=\{{\tt A,C,G,T}\}$,
such that $d({\bf X,x})$ is greater than some threshold value $k$.
\begin{defn}\label{def1}
An \emph{exact nested tandem repeat} is a string of the form
$$
{\mathbf x}^{s_0}{\mathbf X}{\mathbf x}^{s_1}{\mathbf X}\cdots
          {\mathbf X}{\mathbf x}^{s_n},
$$
where $n>1$, $s_{i}\ge1$ for each $0<i<n$, and $s_j \ge 2$ for some $j \in [0,1,\cdots,n]$.
The motif $\bf x$  is called the {\em tandem repeat} and the motif $\bf X$ is the {\em interspersed repeat}.
The concatenations of the tandem repeats ${\bf x}^{s_i}$ alone, and of the interspersed motifs $\bf X$ alone, both form exact tandem repeats.\\
\end{defn}

\textbf{Example:} $\tt {\bf x}= ACGGT$, $\tt {\bf X}= GACCTTTGG$, $n=7$, $s_0=0$, $s_1=3$, $s_2=5$, $s_3=2$, $s_4=4$, $s_5=1$, $s_6=s_7=2$, so
\begin{align*}
{\bf x}^{0}\prod_{i=1}^{7}\mathbf{X}\mathbf{x}^{s_i}
=&\mathbf{XxxxXxxxxxXxxXxxxxXxXxxXxx}\\
=& {\tt GACCTTTGG~ACGGT~ACGGT~ACGGT}\\
 & {\tt GACCTTTGG~ACGGT~ACGGT~ACGGT~ACGGT~ACGGT}\\
 &{\tt GACCTTTGG~ACGGT~ACGGT} \\
 & {\tt GACCTTTGG~ACGGT~ACGGT~ACGGT~ACGGT}\\
 &{\tt GACCTTTGG~ACGGT}\\
 &{\tt GACCTTTGG~ACGGT~ACGGT}\\
 &{\tt GACCTTTGG~ACGGT~ACGGT}.
\end{align*}

In practice we expect any nested tandem repeats occurring in DNA sequences
to be approximate rather than exact. In what follows we will write
$\tilde{\mathbf X}$ to mean an approximate copy of the motif $\mathbf X$,
and $\tilde{\mathbf{x}}^s$ to mean an approximate tandem repeat consisting of
$s$ approximate copies of the motif $\mathbf{x}$.

\begin{defn}
A \emph{($k_1$,$k_2$)-approximate nested tandem repeat} is a string of the form
$$ \tilde{\mathbf x}^{s_0}\tilde{\mathbf X}\tilde{\mathbf x}^{s_1}\tilde{\mathbf X}\cdots \tilde{\mathbf X}\tilde{\mathbf x}^{s_n}, $$
where $n$ and $s_i$ satisfy the same conditions in Definition~\ref{def1}, and  $\tilde{\mathbf x}^{s_0}\tilde{\mathbf
  x}^{s_1}\cdots \tilde{\mathbf x}^{s_n}$ is a $k_1$-approximate tandem
repeat with motif $\mathbf{x}$, and
$\tilde{\mathbf X}\tilde{\mathbf X}\cdots \tilde{\mathbf X}$ is a $k_2$-approximate tandem repeat with
motif $\mathbf{X}$.
\end{defn}
{\bf Examples:}
\begin{itemize}
\item{\bf NTR}:\\ $\tt AGG~AGG~CTCAG~AGG~CTCAG~AGG~AGG~AGG~CTCAG$.\\ The motifs are $\tt AGG, CTCAG.$
\item{\bf $(1,2)-$NTR}:\\ $\tt AG{\underline A}~AGG~CT\underline{TC}G~AGG~CTCAG~AA~AG{\underline A}~AGG~CT{\underline TC}G~AGG\\CTCAG~A{\underline A}G$.\\ The motifs are $\tt AGG, CTCAG$.
\end{itemize}

\section{Related Work}

 Various algorithms have been introduced to find exact tandem repeats. Such algorithms were developed mainly for theoretical purposes, namely, to solve the problem of finding squares in strings \cite{DBLP:journals/tcs/ApostolicoP83,DBLP:journals/ipl/Crochemore81, Kolpakov01findingapproximate, DBLP:journals/jal/MainL84, DBLP:journals/tcs/StoyeG02}.
These algorithms are not easily adapted to finding the approximate tandem repeats that usually occur in DNA.

A number of algorithms,  \cite{DBLP:journals/bioinformatics/DelgrangeR04, DBLP:journals/jcb/LandauSS01} consider motifs differing only by substitutions, using the Hamming distance as a measure of similarity.
Others, e.g.\ \cite{GBenson01151999, DBLP:conf/ismb/HauthJ02, DBLP:journals/jcb/DomanicP07,DBLP:journals/jcb/SagotM98, wexler}, have considered insertions and deletions by using the edit distance. Most of these algorithms have two phases, a scanning phase that locates candidate tandem repeats, and an analysis phase that checks the candidate tandem repeats found during the scanning phase.

The only algorithm designed to look for NTRs is that of Hauth and Joseph (2002), which searches for tandem motifs of length at most six nucleotides.

\section{The Algorithm}

In this section we present the algorithm we have developed to search for nested tandem repeats in a DNA sequence.
The algorithm requires several preset parameters.
These are: $k_1$ and $k_2$ which bound the edit distances from the tandem and interspersed motifs; and the motif length bounds $\min_{t_1},\max_{t_1},\min_{t_2},\max_{t_2}$.
Other input parameters are discussed below.

\paragraph{\bf Search phase}
Our search is confined to seeking NTRs with motifs of length $l_1\in [\min_{t_1},\max_{t_1}]$
and $l_2\in [\min_{t_2},\max_{t_2}].$
A $(k_1,k_2)-$NTR must contain a $k_1-$TR, so we begin by scanning the sequence for approximate tandem repeats.
Several good algorithms, including those of Benson (1999), Wexler {\it et al.} (2005) and Domani\c{c} and Perparata (2007), have been developed to find $k_1-$TRs.
We have chosen to adapt the algorithm of Wexler {\it et al.} (2007),  where the sequence is scanned by two windows $w_1$, $w_2$ of width $w$, a distance $l_1$ apart.
Wexler's algorithm uses a similarity parameter $q$ with default value $q=0.5$, which can be reset by the user.
The user may set the $k_1, k_2$ values, preset with default values
$$k_1=l_1(1-p_m)+\sqrt{l_1(1-p_m)p_m}$$
$$k_2=l_2(1-p_m)+\sqrt{l_2(1-p_m)p_m},$$
following Domani\c{c} and Preparata (2007), with matching probability $p_m$ given the default value $p_m=0.8$.

Once a TR has been found and its full extent determined, the right-most copy of the repeated pattern is taken as the current TR motif $\tt x$, and further approximate copies of $\tt x$ are sought, displaced from the TR up to a distance of $\max_{t_2}$ nucleotides to the right.
If no further approximate copies of $\tt x$ are located, this TR is abandoned, and the TR search continues to the right.
If a displaced approximate copy of $\tt x$ is observed, then both $\tt x$ and the interspersed segment $\tt X$ are recorded in a list, as we have found a candidate NTR.
Further contiguous copies of $\tt x$ are then sought, with the rightmost copy $\tt x$ replacing the previous template motif.

The steps above are repeated with successive motifs $\tt x$ and interspersed segments copied to the list, until no additional copies of the last recorded motif $\tt x$ are found.
This search phase is illustrated in Figure 1.

At this point the algorithm builds consensus patterns for $\tt x$ and $\tt X$ using majority rule. After constructing the two consensus patterns the algorithm moves to the verification phase.

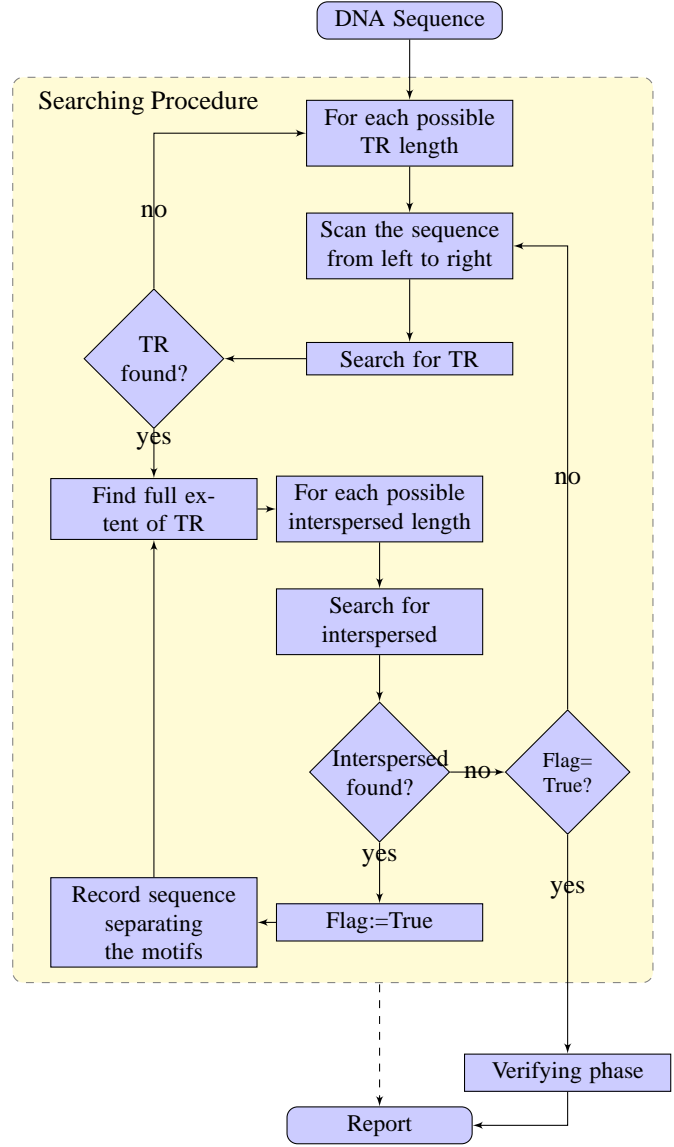
\begin{figure}
\tikzstyle{decision} = [diamond, draw, fill=blue!20,
    text width=4em, text badly centered, node distance=2cm, inner sep=0pt]
\tikzstyle{block} = [rectangle, draw, fill=blue!20,
    text width=8em, text centered, minimum height=1em]
\tikzstyle{start} = [rectangle, draw, fill=blue!20,
    text width=2.5cm, text centered, rounded corners, minimum height=1em]
\tikzstyle{line} = [draw, -latex']
\tikzstyle{cloud} = [draw, ellipse,fill=red!20, node distance=2cm,
    minimum height=1em]
    \tikzstyle{lblock} = [rectangle, draw, fill=blue!20,
    text width=14em, text centered, minimum height=1em]
\tikzstyle{vlblock} = [rectangle, draw, fill=blue!20,
    text width=8cm, text centered, rounded corner, height=18cm]
    \pgfdeclarelayer{background}
\pgfdeclarelayer{foreground}
\pgfsetlayers{background,main,foreground}

\begin{tabular}{c}
\begin{tikzpicture}[node distance = 1.8cm]
    \node [start,scale=0.9] (DNA) {DNA Sequence};
    \node [block, below of=DNA,node distance=1.5cm,scale=0.9] (TRLengthLoop) {For each possible TR length};
    \node [block, below of=TRLengthLoop,node distance=1.5cm,scale=0.9] (SequenceLoop) {Scan the sequence from left to right};
    \node [block, below of=SequenceLoop, node distance=1.5cm,scale=0.9] (CheckSimilarity) {Search for TR};
    \node [decision, left of=CheckSimilarity,node distance=3.4cm,scale=0.9] (IsSimilar) {TR found?};
     \node [block, below of=IsSimilar, node distance=2cm,scale=0.9] (Full) {Find full extent of TR};

    \node [block, right of=Full,node distance=3cm,scale=0.9] (NTRSearch) {For each possible interspersed length};
    \node [block, below of=NTRSearch, node distance=1.5cm,scale=0.9] (CheckSimilarityNTR) {Search for interspersed };
    \node [decision, below of=CheckSimilarityNTR,scale=0.9] (IsSimilarNTR) {Interspersed found?};
    \node [block, below of=IsSimilarNTR,node distance=2cm,scale=0.9] (SetFlag) {Flag:=True};

    \node [block, left of=SetFlag,node distance=3cm,scale=0.9] (Record) {Record sequence separating the motifs};
\node [decision, right of=IsSimilarNTR,node distance=2.5cm,scale=0.8] (Flag) {Flag= True?};
    \node [block, below of=Flag,node distance=4cm,scale=0.9] (Align) {Verifying phase};
    \node[below of=SetFlag,node distance=0.7cm](c){};
    \node [start,below of=c,node distance=2cm,scale=0.9] (Report) {Report};

   \begin{pgfonlayer}{background}

    \path (Full.west |- TRLengthLoop.north)+(-0.5,0.3) node (a) {};
        \path (Record.south -| Flag.east)+(+0.3,-0.2) node (b) {};

    \path[fill=yellow!20,rounded corners, draw=black!50, dashed]
            (a) rectangle (b);
    \end{pgfonlayer}
     \path (TRLengthLoop.west)+(-2.1,0.4) node (Search) {Searching Procedure};

    \path [line,dashed] (c) -- (Report);
    \path [line] (Align) |- (Report);
    \path [line] (DNA) -- (TRLengthLoop);
    \path [line] (TRLengthLoop) -- (SequenceLoop);
    \path [line] (SequenceLoop) -- (CheckSimilarity);
    \path [line] (CheckSimilarity)  -- (IsSimilar);
    \path [line] (IsSimilar) -- node [near start] {yes} (Full);
        \path [line] (Full) -- (NTRSearch);

    \path [line] (IsSimilar) |- node [near start] {no} (TRLengthLoop);

    \path [line] (NTRSearch) -- (CheckSimilarityNTR);
    \path [line] (CheckSimilarityNTR) -- (IsSimilarNTR);

    \path [line] (IsSimilarNTR) -- node [near start] {yes} (SetFlag);
    \path [line] (IsSimilarNTR) -- node  {no} (Flag);
    \path [line] (SetFlag) -- (Record);
    \path [line](Record) -- (Full);
    \path [line] (Flag) -- node [near start] {yes} (Align);
    \path [line] (Flag) |- node [near start] {no} (SequenceLoop);

\end{tikzpicture}
\end{tabular}
\caption{Flowchart of the {\tt NTRFinder} algorithm.}
\label{searchphase}
\end{figure}

\paragraph{\bf Example:}
An example will help illustrate the procedure.
Suppose that $S$ contains an NTR of the form
$$\mathbf{xX_0xxxX_1xxxxxxX_2xxX_3}.$$ The algorithm will scan from
the left until it locates the tandem repeat consisting of three copies
of $\mathbf{x}$ between $\mathbf{X}_0$ and $\mathbf{X}_1$.  It will
then start searching for additional non-adjacent copies of
$\mathbf{x}$ to the right, locating the first copy to the right of
$\mathbf{X}_1$. Having found this it will record the intervening segment $\mathbf{X}_1$, and then continue the tandem repeat search
  from this point until the full extent of the tandem repeat between
  $\mathbf{X}_1$ and $\mathbf{X}_2$ is found.

This procedure is repeated once more, locating the tandem repeat
between $\mathbf{X}_2$ and $\mathbf{X}_3$, recording the
segment $\mathbf{X}_2$, and then searching for further copies to
the right.  At this point no more copies of $\mathbf{x}$ are
found, and the process of verification begins. The segments
$\mathbf{X}_0$, $\mathbf{X}_3$ and the initial copy of $\mathbf{x}$
are found during this stage.

\paragraph{\bf Verification phase:} Each candidate NTR is checked to determine whether it meets the NTR definition.
This is accomplished by aligning the candidate NTR region, together with
a margin on either side of it, against the consensus motifs x and X,
using the nested wrap-around dynamic programming algorithm of
Matroud et al. (2010). This has complexity $O(n|\mathtt x||\mathtt X|)$,  where $n$ is the length of the NTR
region and $|{\tt {\bf x}}|$ and $\tt |{\bf X}|$ are the length of the
tandem motif and the length of the interspersed motif respectively.

\section{Results}

\subsection{Tests on real sequence data}

We have implemented our algorithm and carried out searches for NTRs on
some DNA sequences taken from GenBank. The size ranges used for this search
were $[\min_{t_{1}},\max_{t_{1}}]=[\min_{t_{2}},\max_{t_{2}}]=[2,100]$,
with the parameters $k_1$, $k_2$ and $q$ left set to their default values.
Some NTR regions found by our software are listed in
Table~\ref{sequencesfound}.

\begin{table*}[t]
\label{result}
\resizebox{\textwidth }{!}{
\begin{tabular}{|c|c|c|c|c|}
  \hline
   \multirow{2}{*}{}Species & tandem motif {\bf x}& $|\mathbf{x}|$  & start index &  \# x  \\
        Accession number     & interspersed motif {\bf X}& $|\mathbf{X}|$ & end index  &     \# X   \\ \hline
    \multirow{2}{*}{}\emph{N. sylvestris} & x={ AGGACATGGC}& 10 & 960 & 53   \\
               X76056.1      &X={ CATGGCACGAGTC}& 13& 2,111 & 26 \\ \hline
   \multirow{2}{*}{}\emph{B. juncea} & x={ GGACGTCCCGCGTGCACAGAC}&21 & 1,403 & 51 \\
                X73032.1                &X={ CACAGACGGTCGACCTGGACGACCTGCGTG}&30 &2,605 & 7\\ \hline
   \multirow{2}{*}{}\emph{B. olerecea} & x={ GGACAGTCCTCGTGGGCGAAAATCACCCAC}&30 & 1,256 & 32 \\
                 X60324.1               &X={ GGATAGTCCACGGGAAGGGCCAACGTGCTGATATGCGTACTGAC}&44 & 3,341 &20 \\ \hline
   \multirow{2}{*}{}\emph{B. rapa} & x={ GGATCAGTACAC}&12 & 385 & 20 \\
               S78172.1                 &X={ GTCCACGGGAAGGGCCAACATGCTGATATGTGTAATACACGGACA}&45 & 1,337 & 8 \\ \hline
   \multirow{2}{*}{}\emph{B. campestris} & x={ GGACGTCCTTTGTGTGCTGAC}&21 & 1,558 & 37 \\
                   S78172.1              &X={ GGACACACGGACACACACGGACAGCCACGGGAAGGGCCAGCGTGTGCTGAC}&51 & 2,580 & 8 \\ \hline
  \multirow{2}{*}{}\emph{C. esculenta} & x={ TCGCACAGCCG}&11 & 725 & 94\\
               Not published                 & X={ TTCTGGGCAAAACGGCTGGGTGACGTGCTGAACTGGCCAGCTGGTTCG}&48 &2384 & 12\\ \hline
   \multirow{2}{*}{}\emph{D. melanogaster}& x= { TGCCCCAGT}&9 &   4,215,779& 7 \\
               AE014296.4               & X= { TGCTGCTCGCCTGGC}&15 & 4,215,899 & 6 \\ \hline
   \multirow{2}{*}{}\emph{H. sapien X chromosome}& x= { CT}&2 &   35,471& 360\\
                    AL672277.21         & X= { CACAAGGAGCTGCTCTCCTCCTTTCTTCTGTTGAGACGTGTGTGTGTCTGTCTTT}&55 & 35,711 & 8 \\ \hline
   \multirow{2}{*}{}\emph{H. sapien X chromosome}& x= { GATA}&4 &   111,705& 147 \\
                    AL683871.15            & X= { TGATGGTAATAGATACATACTTAGGTA}&27 & 113,805& 56 \\
  \hline

\end{tabular}
 }

 \caption{Nested tandem repeats found in some sequences from GenBank
and an additional unpublished sequence (\emph{C. esculenta}).}
 \label{sequencesfound}
\end{table*}

\subsection{More complex structures}

In addition to the nested tandem repeats in
Table~\ref{sequencesfound}, {\tt NTRFinder} also
reported an NTR in \emph{Linum usitatissimum} (accession number gi|164684852|gb|EU307117.1|) which on further analysis
by hand turned out to have a more complex structure. The IGS region
of the rDNA of this species contains an NTR with four motifs
interspersed with each other.  The four motifs are $w$={\tt
  GTGCGAAAAT}, $x$={\tt GCGCGCCAGGG}, $y$={\tt GCACCCATAT}, and
$z$={\tt GCGATTTTG} and the structure of the NTR is
$$ \prod_{i=1}^{25}w^{q_i}x^{r_i}z^{s_i}y^{t_i},$$ where $q_i \in \{1,2,3\}$; $r_i \in \{1,2\}$; $s_i \in \{0,1\}$; $t_i \in \{0,1\}$.

\subsection{Running time}

\begin{figure}[b]
\label{performance}

\begin{tikzpicture}
    \begin{loglogaxis}[xlabel= Sequence length (bp), ylabel= Time (s)]
    \addplot+[only marks,mark=*,blue] plot coordinates {
        (2646,0.5)
        (3451,0.6)
        (5488,1.5)
        (14530,6)
        (17795,3)
        (30177,35)
        (91260,17)
        (163080,31)
        (1409992,256)

    };

    \end{loglogaxis}
    \end{tikzpicture}

 \caption{Running time of NTRFinder (on a Pentium Dual core T4300 2.1 GHz) plotted against segment length on a log-log scale. The search was performed on segments of different lengths, with the minimum and maximum tandem repeat lengths set to 8 and 50 respectively. The distribution suggests the running time is approximately linear with sequence length.}
 \label{runningtime}
\end{figure}
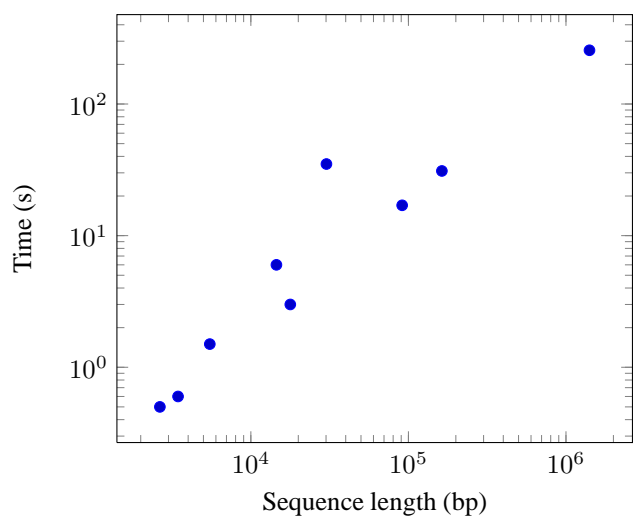

The running time for {\tt NTRFinder} searching some sequences from
GenBank is shown in Figure~\ref{runningtime}. It can be seen that the
run time is approximately linear in the length of the sequence.
However, it must be noted that the run time depends not only on the
length of the input sequence, but also on the number of tandem and
nested tandem repeats found in the sequence. The program spends most of the
time verifying any tandem repeats found.

\section{Discussion}
In the last decade a number of software tools to find tandem repeats
have been introduced; however, little work exists on more complex
repetitive structures such as nested tandem repeats. The problem of
finding nested tandem repeats is addressed in this study.  The
motivation for our study is the potential use of NTRs as a
marker for genetic studies of populations and of species.

We have done some analysis on the nested tandem repeat in the
intergenic spacer region in {\it C. esculenta} (taro), noting some
variation in the NTRs derived from domesticated varieties sourced from
New Zealand, Australia and Japan. Further varieties are currently
being analysed.  By considering some edit operations such as deletion,
mutation, and duplication we can align the nested tandem repeat
regions of each pair of sequences. The alignment score can then be
considered as a measure of distance between both sequences.  In
particular they appear to share some common inferred histories of the
development of the NTRs from a simpler structure of two motifs.  The
edit operations appear to be occurring on a 1,000 year timescale, so
this analysis offers the potential to date the prehistory of the early
agriculture of this ancient staple food crop.

\section{Conclusion}
The nested tandem repeat structure is a complex structure that
requires further analysis and study.  The number of copy variants in
the NTR region and the relationships between these copies might
suggest a tandem repeat generation mechanism.  In this paper, we have
introduced a new algorithm to find nested tandem repeats. The first phase of the algorithm has $O(n(\max_{t_1})(\max_{t_2}))$ time complexity, while the second phase (the alignment) needs $O(n(\max_{t_1})(\max_{t_2}))$ space and time, where $n$ is the length of the NTR region, and $\max_{t_1}, \max_{t_2}$ are the maximum allowed lengths of the tandem and interspersed motifs.

\section*{Acknowledgements}
Andrew Clarke and Peter Matthews, for providing data and useful background about Taro, and Hussain Matawa, for assisting in the development of the program interface.

\paragraph{Funding: This project was partially funded by the Allan Wilson Centre for Molecular Ecology and Evolution.}

\bibliography{biblio}

\end{document}